\documentclass[prb,twocolumn,showpacs,preprintnumbers,superscriptaddress,amsmath,amssymb]{revtex4}
\usepackage{graphicx}
\usepackage{dcolumn}
\usepackage{epsfig}

\begin{document}

\title{High pressure effects on single crystal electron-doped Pr$_{2-x}$Ce$_{x}$CuO$_{4}$}

\author{C. R. Rotundu}
\email[On leave from Lawrence Berkeley National Laboratory; \\corresponding author: CostelRRotundu@gmail.com\\]{}
\affiliation{Center for Nanophysics \& Advanced Materials and Department of Physics, University of Maryland, College Park, MD 20742, USA}
\affiliation{Materials Sciences Division, Lawrence Berkeley National Laboratory, Berkeley, CA 94720, USA}
\author{V. V. Struzhkin}
\affiliation{Geophysical Laboratory, Carnegie Institution of Washington, Washington, DC 20015, USA}
\author{M. S. Somayazulu}
\affiliation{Geophysical Laboratory, Carnegie Institution of Washington, Washington, DC 20015, USA}
\author{S. Sinogeikin}
\affiliation{HPCAT, Geophysical Laboratory, Carnegie Institution of Washington, Advanced Photon Source, Argonne National Laboratory, Argonne, Illinois 60439, USA}
\author{Russell J. Hemley}
\affiliation{Geophysical Laboratory, Carnegie Institution of Washington, Washington, DC 20015, USA}
\author{R. L. Greene}
\affiliation{Center for Nanophysics \& Advanced Materials and Department of Physics, University of Maryland, College Park, MD 20742, USA}
\date{\today}

\begin{abstract}

We present high pressure diamond anvil cell synchrotron X-ray, resistivity, and ac-susceptibility measurements on the electron-doped cuprate Pr$_{2-x}$Ce$_{x}$CuO$_{4}$ to much higher pressures than previously reported. At 2.72 GPa between 88 and 98$\%$ of the superconducting T$^\prime$ phase of the optimally doped Pr$_{1.85}$Ce$_{0.15}$CuO$_{4}$ transforms into the insulating phase T. With application of pressure, the T phase becomes more insulating, so we present here what may be the first example of electron-doping in the T structure. The results have implications for the search for ambipolar high-T$_{c}$ cuprate superconductors. The T$_{c}$ of the remaining 2-12$\%$ T$^\prime$ phase is suppressed continuously from 22 K to 18.5 K at about 14 GPa. Remarkably, the T$_{c}$ of the overdoped Pr$_{1.83}$Ce$_{0.17}$CuO$_{4}$ remains practically unchanged even at 32 GPa.

\end{abstract}


\maketitle

\section{Introduction}

Although hole-doped cuprates are the most studied class of high-T$_{c}$ materials, attention has been drawn recently to the electron-doped cuprates \cite{Greene1,Greene2} in the effort to achieve a unified understanding of the high temperature superconducting mechanism in cuprates. High pressure experiments are important for understanding the superconductivity and to help identify ways for increasing T$_{c}$. Experiments on hole-doped cuprates showed an increase of T$_{c}$ when pressure is applied, with the record belonging to HgBa$_{2}$Ca$_{2}$Cu$_{3}$O$_{8+\delta}$ \cite{Gao} for which the T$_{c}$ is enhanced from 133 K to 164 K when compressed to 30 GPa. Pressures up to 2.5 GPa showed no (or extremely small) changes in structural \cite{Kamiyama} and other physical properties \cite{Murayama,Markert} of electron-doped cuprates. We present here a high pressure study, to pressures higher than previously reported, of the structural and other physical properties of single crystals of electron-doped Pr$_{2-x}$Ce$_{x}$CuO$_{4}$. We explain the close relation between the structural and superconducting properties. To our best knowledge, there are no high pressure ($>$ 2.5 GPa) studies of the superconducting properties of electron-doped cuprates, except for the resistivity study on polycrystalline Ln$_{1.85}$Ce$_{0.15}$CuO$_{4-y}$ to 10 GPa by J. Beille $\emph{et al.}$ \cite{Beille}.

\section{Experimental Methods}

Single crystals of Pr$_{2-x}$Ce$_{x}$CuO$_{4}$, x=0.15 (optimally-doped) and 0.17 (over-doped) were synthesized via a flux method refined by Peng $\emph{et al.}$ \cite{Peng}. The x=0.15 crystal had T$_{c}$ = 21 K under normal pressure conditions as determined from magnetization measurement in 20 Oe, in agreement with literature values \cite{Peng}. Diamond anvil cell (DAC) high pressure resistivity measurements were run on a small sample of approximately 40 $\times$ 40 $\times$ 10 $\mu$m$^{3}$ cleaved from a few mm size crystal. The measurements were performed using a standard four-probe Van der Pauw configuration, and the schematic of the setup is shown in Fig. 1a.

Pressure was achieved using a lever-arm system with two 300 $\mu$m culet diamonds mounted on tungsten carbide supports. On the culet of one of the diamonds four radial platinum-based polymer conductive leads were deposited using focused ion beam (FIB) lithography \cite{Melngailis}.  The inner end of these leads passed over the sample, thereby assuring electrical contact and mechanical attachment of the sample to the diamond (Fig. 1b). A stainless steel gasket was indented first to 40 $\mu$m thickness and a centered $\sim$ 100 $\mu$m hole in the indentation was drilled. Cubic boron nitride (BN) powder was indented in the hole and on the conical side of the gasket, creating a thin insulating layer. Four electrodes 5 $\mu$m made of thin platinum foil were indented in a radial position to assure electrical contact with the FIB depositions (Fig. 1d). The BN was drilled in the center to match the gaskets hole, and the space created formed the sample chamber.  Before closing the DAC, ruby spheres were placed next to the sample for in situ pressure determination based on a calibrated fluorescence shift \cite{Mao}. Finally, the DAC was closed in pre-compressed Ne gas at about 0.2 GPa, which served as a pressure-transmitting medium. During the course of the experiment, it was essential that the diamond anvils do not press directly on the sample.  Thus, the thicknesses of the indented gasket and sample were 40 $\mu$m and 10 $\mu$m, respectively. Measurements were stopped on compression when the gasket thinned down such that the sample was in direct contact with the diamonds. We note that the material under study cannot withstand an uniaxial pressure larger than 0.5 GPa \cite{Kaga}. In the present resistivity measurement the pressure corresponding to gasket collapse was larger than 43 GPa.

DAC magnetic ac-susceptibility measurements were carried out using a double-frequency modulation method with details given elsewhere \cite{Struzhkin,Gregoryanz,XJChen}. The DAC consisted of two pairs of diamonds inside of a larger primary coil, a secondary signal coil (encircling the pair of the diamonds with the sample), and secondary compensating coil - identical to the secondary signal coil but with no sample. The gasket, made of a NiCrAl non-magnetic alloy, was pre-indented by the two pairs of diamonds and then drilled the center of the indentations. The crystal was cut to approximately 50 $\times$ 50 $\times$ 20 $\mu$m$^{3}$ and placed inside of one of the drilled holes, and the other was left blank intentionally.

High pressure X-ray diffraction was performed on powder from crushed crystals at HPCAT (Sector 16) at the Advanced Photon Source (APS), Argonne National Laboratory. In these experiments, a DAC with 400 $\mu$m culet diamonds with a sample chamber having a 50 m m diameter was used. As for the resistivity
and ac-susceptibility measurements, Ne was used as the pressure medium. Ne provides an quasihydrostatic environment for pressures up to about 15 GPa; bove this value, the pressure gradients remain very small: at 50 GPa the standard deviation of pressure is less than 1$\%$ \cite{Klotz}.

\begin{figure}[h]
\begin{center}\leavevmode
\includegraphics[width=0.96\linewidth]{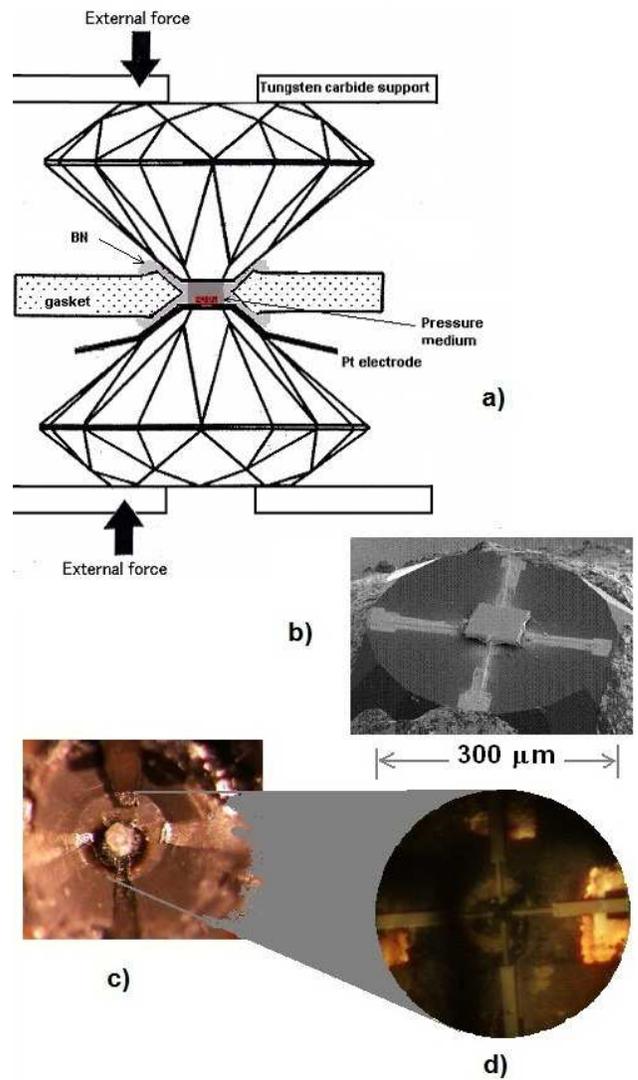}
\caption {a) Schematic of the diamond anvil cell resistivity high pressure setup. b) The diamond culet with the platinum-based polymer contacts on the sample. c) Indented Boron Nitride with platinum-foil made leads. The center hole becomes the sample space. d) View of the sample with the electrical contacts after DAC assembly was closed under compressed neon gas and ready for the experiment. (details in the text)}
\label{fig2}\end{center}\end{figure}

\section{Results and Discussion}

\begin{figure}[h]
\begin{center}\leavevmode
\includegraphics[width=1.02\linewidth,bb=20 20 380 460]{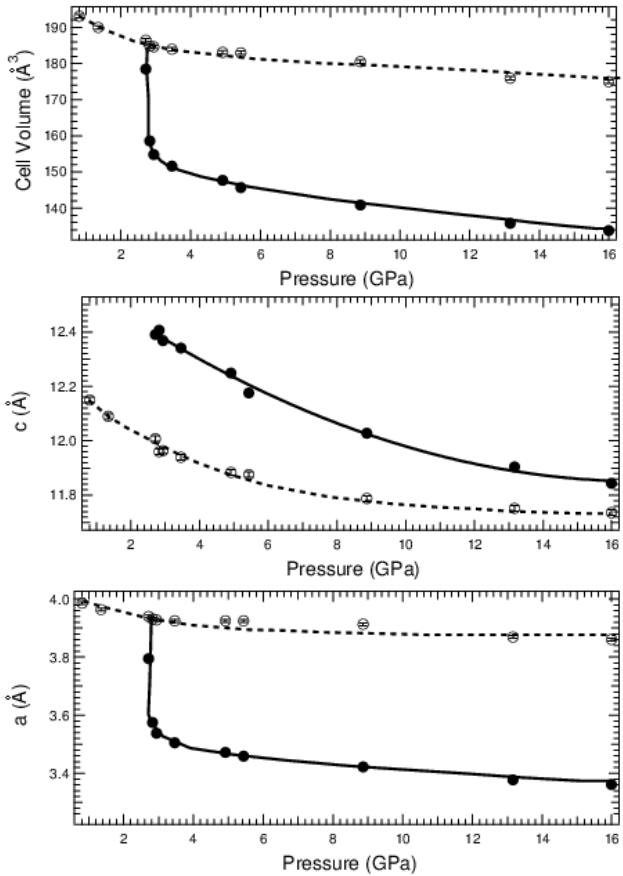}
\caption {Lattice parameters versus pressure of Pr$_{1.85}$Ce$_{0.15}$CuO$_{4}$. From top to bottom: Volume cell versus pressure (P), $c$ axis versus $P$, and $a$ axis versus $P$. Solid symbols are for the T phase,
hollow for the T$^\prime$ and the lines are a guide to the eye. At 2.72 GPa between 88 and 98$\%$ of the T$^\prime$ phase transforms to the T phase.}
\label{fig2}\end{center}\end{figure}

An early high pressure X-ray (to 0.6 GPa) study by Kamiyama $\emph{et al.}$ \cite{Kamiyama} showed a very small but clear decrease of the lattice parameters with pressure of the undoped Nd$_{2}$CuO$_{4}$ and optimally-doped Nd$_{1.835}$Ce$_{0.165}$CuO$_{4}$. Higher pressure experiments showed that in the parent Nd$_{2}$CuO$_{4}$ a T$^\prime$ \cite{Tprime} to T structural transition takes place at 21.5 GPa \cite{Wilhelm2}, but the transition is found to take place at 15.1 GPa in the parent Pr$_{2}$CuO$_{4}$ \cite{Wilhelm1}. To our best knowledge we are the first to show X-ray data for an electron-doped cuprate to a pressure higher than 0.6 GPa. Figure 2 show the lattice parameters $a$ and $c$ and volume cell versus pressure of the optimally-doped Pr$_{1.85}$Ce$_{0.15}$CuO$_{4}$. Our first pressure data point is 0.8 GPa. Interestingly, the T$^\prime$ to T transition takes place at a much lower pressure, 2.72 GPa, when 88-98$\%$ of the T$^\prime$ phase transforms to the T phase (Fig. 3). This is of interest because for the case of the undoped Pr$_{2}$CuO$_{4}$ at 37.2 GPa there is still 50$\%$ of the T$^\prime$ phase surviving \cite{Wilhelm1}. While we believe the differences are mostly intrinsic, the different pressure media used (N2 gas in Wilhelm $\emph{et al.}$ \cite{Wilhelm1,Wilhelm2} vs. the more hydrostatic Ne gas in the present study) may have a sizable influence. The standard deviation of pressure is about 3-4$\%$ in N2 gas at 25 GPa, while for the case of Ne is less than 1$\%$ even at 50 GPa \cite{Klotz}. One question remaining to be addressed is up to what pressure the phase T$^\prime$ coexists with T in the optimally doped cuprate. In Pr$_{2}$CuO$_{4}$ both T$^\prime$ and T are present in a 50$\%$ ratio up to 37.2 GPa. In Nd$_{2}$CuO$_{4}$ the phases coexist for the [21.5, 29.5] GPa pressure interval \cite{Wilhelm2} and shorten further to [11.4, 15] GPa for LaNdCuO$_{4}$ \cite{Wilhelm1}. Upon applying pressure, the lattice constants of the T$^\prime$ phase of Pr$_{1.85}$Ce$_{0.15}$CuO$_{4}$ are continuously suppressed through the phase transition as seen in Fig. 2. One other observation is that 16 GPa pressure produces a more drastic shrinkage of the lattice parameters of the T$^\prime$ phase than a 23$\%$ Ce substitution of Pr \cite{Maiser}. In fact, the 23$\%$ Ce doping (maximal solubility \cite{Peng}) produces lattices changes equivalent to about 2 GPa of pressure.

\begin{figure}[h]
\begin{center}\leavevmode
\includegraphics[width=0.98\linewidth]{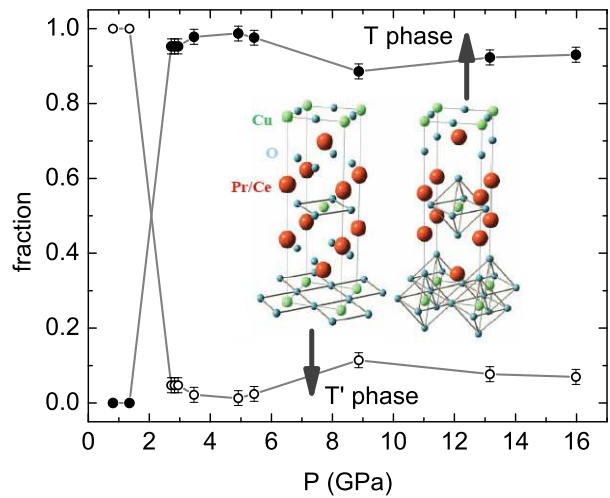}
\caption {Fractions of phases T$^\prime$ and T versus pressure for Pr$_{1.85}$Ce$_{0.15}$CuO$_{4}$. Inset shows representations of both T$^\prime$ and T structures \cite{powdercell}.}
\label{fig2}\end{center}\end{figure}

\begin{figure}[h]
\begin{center}\leavevmode
\includegraphics[width=0.98\linewidth]{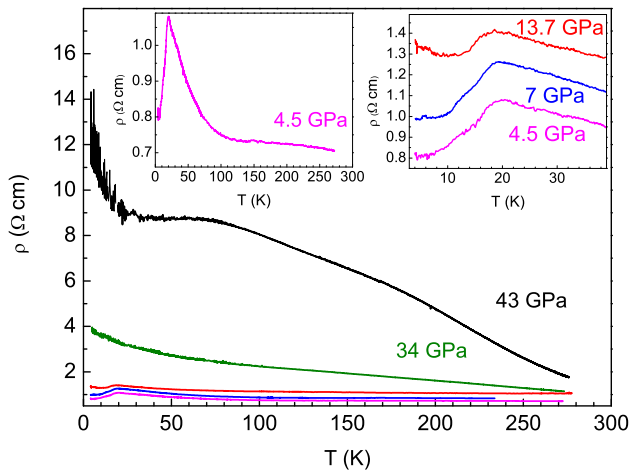}
\caption {Resistivity $\rho$ versus temperature of Pr$_{1.85}$Ce$_{0.15}$CuO$_{4}$ at 4.5, 7, 13.7, 34, and 43 GPa.}
\label{fig2}\end{center}\end{figure}

\begin{figure}[h]
\begin{center}\leavevmode
\includegraphics[width=1.04\linewidth]{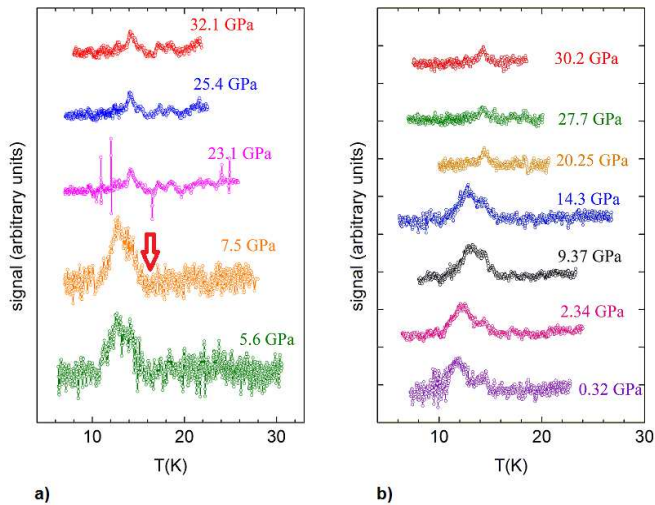}
\caption {Real component of the ac-susceptibility versus temperature of the over-doped Pr$_{1.83}$Ce$_{0.17}$CuO$_{4}$ at various pressures during a) compression and b) decompression. Data are vertically displaced for clarity. The arrow (shown for the 7.5 GPa data) points to the T$_{c}$.}
\label{fig2}\end{center}\end{figure}

Figure 4 shows resistivity versus temperature of Pr$_{1.85}$Ce$_{0.15}$CuO$_{4}$ at 4.5, 7.0, 13.7, 34, and 43 GPa. The attempt to compress the DAC to the next higher pressure resulted in the collapse of the metallic gasket and therefore end of the resistivity experiment. At relatively low pressure the resistivity versus temperature curves show what resembles a superconducting transition (but with non-zero resistivity below T$_{c}$) and enhancement of resistivity close to T$_{c}$ (left inset of Fig. 4). This enhancement of the resistivity near T$_{c}$, is due in part to a slight inclination of the sample cleaved face from the CuO$_{2}$ planes \cite{Suzuki} and in part to granular effects within the crystal \cite{Crusellas1}. It is unlikely that this resistivity enhancement near T$_{c}$ is due to inhomogeneities in Ce doping (as proposed by Klimczuk $\emph{et al.}$ \cite{Klimczuk}) given that the enhancement in the x=0.15 crystal (as seen in resistivity data at 4.5 GPa) is measured on a 10 $\mu$m thickness crystal while the inhomogeneities in Ce were found to appear more in crystals of thickness greater than 300 $\mu$m \cite{Matsuda,Skelton}.

The non-zero resistivity below T$_{c}$ for pressures greater than the 2.72 GPa of the T$^\prime$ $\longrightarrow$ T transition can be explained based on the 88-98$\%$ insulating T phase. This is consistent with the magnitude of resistivity at 4.5 GPa that is of an order of a fraction of a $\Omega$cm, while typical resistivity above T$_{c}$ at normal pressure (where the material is in T$^\prime$ phase) is a fraction of m$\Omega$cm \cite{Peng}. The same mechanism most likely is responsible for the high pressure non-zero resistivity data of Beille \emph{et al.} \cite{Beille} below the superconducting transition in Ln$_{1.85}$Ce$_{0.15}$CuO$_{4-y}$ (Ln = Nd, Sm, Eu), although no high pressure X-ray are available for these compositions.

T$_{c}$ is suppressed by pressure and at 34 GPa we cannot detect any sign of superconducting transition in the resistivity data, and the shape of the resistivity versus temperature curves are consistent with an insulating behavior. At higher pressure (43 GPa), the resistivity versus temperature curve show two broad peaks. These mysterious features are perhaps due to the complicated effect of the pressure on the magnetic ordering (spin orientation) \cite{Katano,Katano2}, and a better understanding of this will require a careful high pressure neutron scattering study.

Figure 5 shows DAC ac-susceptibility data (real component) for the over-doped Pr$_{1.83}$Ce$_{0.17}$CuO$_{4}$ at compression (Fig. 5a) and decompression (Fig. 5b), with maximum pressure of 32.1 GPa. The arrow points to the T$_{c}$. The shape of susceptibility data and how T$_{c}$ is determined when a double-frequency modulation technique is used have been discussed in detail in a review paper by Struzhkin $\emph{et al.}$ \cite{Struzhkin} and are based on the Hao-Clemm theory for reversible magnetization in type II superconductors \cite{Hao-Clemm}. Basically, T$_{c}$ in the ac-susceptibility data for Pr$_{1.83}$Ce$_{0.17}$CuO$_{4}$ is marked by the higher temperature ``end'' of the peak as shown in Fig. 5. The extremely sensitive ac-susceptibility proved to be an excellent probe for detecting and monitoring the evolution of T$_{c}$ versus pressure given the small fraction of the T$^\prime$ superconducting phase beyond the structural transition. Remarkably, for the over-doped Pr$_{1.83}$Ce$_{0.17}$CuO$_{4}$, T$_{c}$ remains unaltered all the way up to 32.1 GPa.

Finally, the phase diagram T$_{c}$ versus pressure is drawn in Fig. 6, for both Pr$_{1.85}$Ce$_{0.15}$CuO$_{4}$ and Pr$_{1.83}$Ce$_{0.17}$CuO$_{4}$. T$_{c}$ for the optimally-doped Pr$_{1.85}$Ce$_{0.15}$CuO$_{4}$ is given by the temperature at the peak of resistivity versus T, and T$_{c}$ for the over-doped Pr$_{1.83}$Ce$_{0.17}$CuO$_{4}$ from the ac-susceptibility versus T as described earlier. We also included in the superconducting phase diagram T$_{c}$ versus pressure for [0-2] GPa as determined from resistivity measurements by Crusellas $\emph{et al.}$ \cite{Crusellas2} on an optimally-doped PCCO crystal. It should be noted here that the high-pressure data points of Crusellas $\emph{et al.}$ \cite{Crusellas2} were obtained from data using 1:1 isoamyl and $n$-pentane alcohol, that is a completely different pressure medium than the neon gas used in the present study. Therefore, lower pressures resistivity measurements using the same Ne gas pressure media will be needed to settle if T$_{c}$ is monotonically suppressed with applying pressure or if that beyond 2.7 GPa (corresponding to the T$^\prime$ $\longrightarrow$ T transition) T$_{c}$ is suppressed at a higher rate. Regardless, the rate of suppression of T$_{c}$ for the optimally-doped sample decreases beyond a pressure that is somewhere between 7 and 13 GPa showing a ``saturation'' to certain T$_{c}$.

\begin{figure}[h]
\begin{center}\leavevmode
\includegraphics[width=0.94\linewidth]{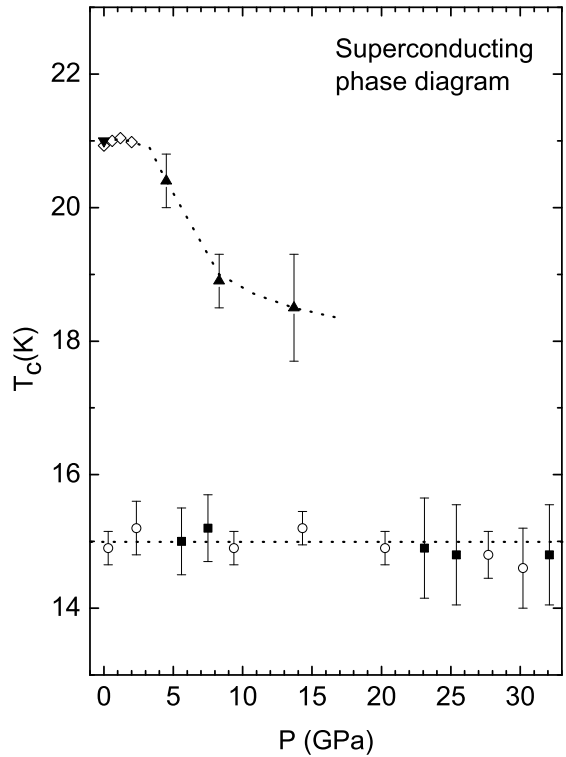}
\caption {T$_{c}$ versus pressure for Pr$_{1.85}$Ce$_{0.15}$CuO$_{4}$ ($\blacktriangledown$ the onset of diamagnetism from $\chi$ under normal pressure, $\diamond$ adapted from resistivity on crystal by Crusellas $\emph{et al.}$ \cite{Crusellas2}, $\blacktriangle$ $\rho$ during compression) and Pr$_{1.83}$Ce$_{0.17}$CuO$_{4}$ ($\bullet$ $\chi$ during compression, $\circ$ $\chi$ during decompression). The dotted lines are a guide to the eye.}
\label{fig2}\end{center}\end{figure}

Lastly we discuss the significance of the resistivity of the optimally-doped (Fig. 4) ``moving'' into a more insulating regime with application of pressure in the context of search for ambipolar \cite{Ando1,Ando2,Ando3} high-T$_{c}$ cuprate superconductors. One such example of an ambipolar high-T$_{c}$ cuprate superconductors has been reported recently by K. Segawa and Y. Ando \cite{Ando1}. They reported successful doping of $n$-type carriers by La substitution for Ba in YBa$_{2}$Cu$_{3}$O$_{y}$, such that Y$_{0.38}$La$_{0.62}$Ba$_{1.74}$La$_{0.26}$Cu$_{3}$O$_{y}$ is 2$\%$ electron-doped. It has been known for a long time that the T-structure can be only easily hole-doped, while the T$^\prime$-structure can be easily only electron-doped \cite{Burns}. In the present study, since 88-98$\%$ of the normal pressure T$^\prime$ phase (that is electron-doped) transforms into the T phase, it is natural to assume that excess electrons were doped in the T phase. We believe the significance of the resistivity of the T phase becoming more insulating with application of pressure is that we successfully doped for the first time $n$-type carriers in the T structure. From the X-ray data it can be seen that T structure is stable up to 16 GPa, so, one question is if the structure is stable at much higher pressures.

\section{Summary}

We studied the evolution of superconductivity and structure (and the relationship between) with pressure of electron-doped Pr$_{2-x}$Ce$_{x}$CuO$_{4}$. At 2.72 GPa between 88 and 98$\%$ of the superconducting T$^\prime$ phase of the optimally-doped Pr$_{1.85}$Ce$_{0.15}$CuO$_{4}$ transforms into the insulating T phase. T$_{c}$ of the remaining 2-12$\%$ T$^\prime$ phase is suppressed from 22 K to 18.5 K at a pressure of about 14 GPa. The non-zero resistivity below T$_{c}$ can be explained based on the 88-98$\%$ insulating T phase for pressures beyond 2.72 GPa. This is in accord with the high magnitude (order of $\Omega$cm) of resistivity at 4.5 GPa, while the typical resistivity above T$_{c}$ at normal pressure (at which the material is in T$^\prime$ phase) is a fraction of m$\Omega$cm \cite{Peng}. T$_{c}$ of the over-doped Pr$_{1.83}$Ce$_{0.17}$CuO$_{4}$ remains practically unchanged even at 32.1 GPa.

One very interesting and surprising result is that with application of pressure, the T phase becomes more insulating, and so we present here the first example of electron-doping in the T structure. One of the most important questions is if by applying even larger pressure the T phase can be driven to electron-doped superconductivity. Most certainly the present study will spark interest and further experiments on the affect of high pressure on the electron-doped cuprates.

\begin{acknowledgments}
We thank P. Fournier and S. Uchida for useful discussions. The FIB deposition contacts were done at Institute for Research in Electronics and Applied Physics, University of Maryland. The work was supported by the State of Maryland and the NSF through grant DMR-1104256 (C.R.R. and R.L.G.), and DOE through DE-FG02-02ER45955 (V.V.S). X-ray diffraction was performed at HPCAT (Sector 16), Advanced Photon Source (APS), Argonne National Laboratory. HPCAT operations were supported by CIW, CDAC, UNLV and LLNL through funding from DOE-NNSA and DOE-BES, with partial instrumentation funding by NSF. APS was supported by DOE-BES, under Contract No. DE-AC02-06CH11357.
\end{acknowledgments}


\begin{thebibliography}{34}
\bibitem{Tprime}
T$^\prime$ - structure is $I$4/$mmm$, Nd$_{2}$CuO$_{4}$-type and T - structure is $I$4/$mmm$, K$_{2}$NiF$_{4}$ - type (see the inset of Fig. 3).
\end{thebibliography}
\end{document}